\documentclass[preprintnumbers, prd, showpacs, floatfix,twocolumn,
preprintnumbers, letterpaper, superscriptaddress]{revtex4}
\usepackage{amsfonts}
\usepackage{amsmath}
\usepackage{amssymb,epsf}
\usepackage{latexsym}
\usepackage{graphicx,epsfig}
\usepackage{amssymb}
\usepackage{subfigure}
\usepackage[colorlinks=true,citecolor=blue,linkcolor=blue,urlcolor=black]{hyperref}
\usepackage{epstopdf}

\newenvironment{itemizeReduced}{
\begin{list}{\labelitemi}{\leftmargin=1em}
\setlength{\itemsep}{1pt}
\setlength{\parskip}{0pt}
\setlength{\parsep}{0pt}}{\end{list}
}
\graphicspath{{Images/}}

\begin{document}

\title{Critical behavior of charged dilaton black holes in AdS space}
\author{Amin Dehyadegari}
\affiliation{Physics Department and Biruni Observatory, Shiraz University, Shiraz 71454,
Iran}
\author{Ahmad Sheykhi}
\email{asheykhi@shirazu.ac.ir}
\affiliation{Physics Department and Biruni Observatory, Shiraz University, Shiraz 71454,
Iran}
\affiliation{Max-Planck-Institute for Gravitational Physics (Albert-Einstein-Institute),
14476 Potsdam, Germany}

\begin{abstract}
We revisit critical behaviour and phase structure of charged anti-deSitter
(AdS) dilaton black holes for arbitrary values of dilaton coupling $\alpha$,
and realize several novel phase behaviour for this system. We adopt the
viewpoint that cosmological constant (pressure) is fixed and treat the
charge of the black hole as a thermodynamical variable. We study critical
behaviour and phase structure by analyzing the phase diagrams in $T-S$ and $%
q-T$ planes. We numerically derive the critical point in terms of $\alpha$
and observe that for $\alpha =1$ and $\alpha \geq \sqrt{3}$, the system does
not admit any critical point, while for $0<\alpha <1$, the critical
quantities are not significantly affected by $\alpha$. We find that unstable
behavior of the Gibbs free energy for $q<q_{c}$ exhibits a \textit{first
order} (discontinuous) phase transition between small and large black holes
for $0\leq\alpha<1$. For $1<\alpha <\sqrt{3}$ and $q>q_{c}$,
however, a novel first order phase transition occurs between small and large
black hole, which has not been observed in the previous studies on phase
transition of charged AdS black holes.
\end{abstract}

\maketitle


\section{Introduction \label{Intro}}

Phase transition is certainly one of the most intriguing and interesting
phenomena in the thermodynamic description of black holes which may shed
some light on the nature of quantum gravity. In particular, the
investigations on the black hole phase structure/transition in an
asymptotically AdS spacetime have received considerable attentions in the
past years. This is mainly due to the remarkable duality between gravity in
an $(n+1)$-dimensional AdS spacetime and the conformal field theory living
on the boundary of its $n$-dimensional spacetime, (AdS/CFT) correspondence.
Perhaps, one of the earliest studies in this direction was done by Hawking
and Page \cite{HP}, who disclosed that there is indeed a first order phase
transition, latter named Hawking-Page phase transition, between the thermal
radiation and the stable large schwarzschild black hole with spherical
horizon in the background of AdS spacetime. Later, Witten discovered \cite%
{Witten} that this phase transition can be interpreted in the AdS/CFT
duality as the confinement/deconfinement phase transition in the strongly
coupled gauge theory. Recently, it has been shown that, for charged AdS
black hole, a second and first order phase transition occurs between small
and large black holes in an extended phase space which resembles the
liquid-gas phase transition in the usual Van der Waals liquid-gas system 
\cite{Dolan,PV}. In an extended thermodynamic phase space, the cosmological
constant (AdS length) is considered as a thermodynamic pressure which can
vary, and its corresponding conjugate quantity is the thermodynamic volume
of the black hole. Taking into account the variation of pressure in the
first law, one observes that the mass of AdS black hole is equivalent to the
enthalpy \cite{enthalpy}. In the recent years, thermodynamic phase
transitions in an extended phase space have been explored for various types
of black holes in AdS space (see Refs. \cite%
{Hendi,Sherkat,Sherkat1,Rabin,Kamrani,DSD1,DSD2,DSDH,
Majhi,superfluidBH,Dehghani,BIRPT,KerrRPT,DRPT,MicroscopicRPT,RupPRL,ZDSH}
and references therein). The studies on phase structure of charged AdS
dilaton black hole have been carried out from both thermal and dynamical
point of view \cite{PHQNM}, where the cosmological constant appears as a
thermodynamical variable. They found that for small dilaton coupling, $%
\alpha \approx 0.01$, the system resembles the Van der Waals fluid
behaviour, while for $\alpha >1$, the $P-v$ diagram of the system deviates
and new phenomena beyond the Van der Waals liquid-gas-like appears \cite%
{PHQNM}. Recently, a novel phase behaviour represents a small/large black
hole \textit{zeroth-order} phase transition, in an extended phase space with
varying cosmological constant, has been observed for charged dilaton black
holes, where the geometry of spacetime is not asymptotically AdS \cite{NAAA}.

Another possible approach to study thermodynamic phase structure of black
hole is to consider the variation of electric charge of the black hole,
while the cosmological constant (AdS length) is kept fixed. With regard to
this perspective, the critical behavior and phase transition of charged AdS
black holes were investigated in a fixed AdS geometry, indicating that it
exhibits the small/large black hole phase transition of Van der Waals type 
\cite{VDW1,AAA}. Interestingly enough, it has been realized in \cite{AAA}
that the phase transition of charged AdS black hole can occur in $Q^2-\Psi$
plane, where $\Psi =1/2r_{+}$ is the conjugate of $Q^2$, without extending
the phase space. Indeed, in this alternative perspective, the relevant
response function clearly signifies the stable and unstable region. Also
there still exist a deep analogy between critical phenomena and critical
exponents of the system with those of Van der Waals liquid-gas system \cite%
{AAA}. The advantages of this new approach is that one do not need to extend
the phase space by treating the cosmological constant as a thermodynamical
variable which may physically not make sense \cite{AAA}. It has been
confirmed that this new approach also works in other gravity theories \cite%
{DS,Homa} as well as in higher spacetime dimensions \cite{Arab}. Recently,
the universality class and critical properties of any AdS black hole,
independent of spacetime metric, via an alternative phase space has also
been explored \cite{AMajhi}. It was shown that the values of critical
exponents for generic black hole are the same with the Van der Waals fluid
system \cite{AMajhi}. For the Born-Infeld AdS black holed in four
dimensional spacetime, we analytically calculated the critical point by
studying the behavior of specific heat in a fixed AdS geometry \cite%
{ASheykhi}. Furthermore, the interesting reentrant phase transition of
Born-Infeld AdS black hole has been investigated in the thermodynamic phase
space \cite{ASheykhi}. The structure of charged black holes has been studied
by employing the Ruppeiner geometry \cite{Kord}, in which the charge is
allowed to vary.

In this paper, we study thermodynamic properties of charged AdS dilaton
black hole in ($3+1$)-dimensional spacetime. Our work differs from \cite%
{PHQNM}, in that we keep the cosmological constant as a fixed quantity and
treat the charge of the black hole as the thermodynamic variable, while the
authors of \cite{PHQNM} extended the phase space by treating cosmological
constant as a variable. Besides, we analyze the phase structure in $T-S$ and 
$q-T$ planes and observe a novel first order phase transition, which has not
been reppored in the previous investigations on phase transition of charged
AdS black holes. As we shall see, the behavior of black hole temperature
crucially depends on the dilaton-electromagnetic coupling constant ($\alpha $%
) for the small horizon radius. When $\alpha =1$ and $\alpha \geq \sqrt{3}$,
we cannot realize any critical point in the system. Besides, for $0\leq
\alpha <1$, we observe a small/large first order phase transition for $%
q<q_{c}$, while similar behaviour is seen for $1<\alpha <\sqrt{3}$ provided $%
q>q_{c}$, where $q_{c}$ is the critical charge of the black hole.

This paper is structured in the following manner. In Sec. II, a brief
overview on thermodynamics of the four-dimensional charged dilaton black
hole in the AdS background is given. In Sec. \ref{Critical}, we investigate
critical behavior of charged dilaton AdS black holes by studying the
specific heat at constant electric charge in $T$-$S$ plane. In Sec. \ref%
{Gibbs}, we use the Gibbs free energy to determine the possible phase
transition in the system. Finally, we summarize the main results Sec. \ref%
{Summary}.


\section{Charged dilaton black holes in AdS space \label{Review}}

We start with a brief review on charged AdS black holes in dilaton gravity
and calculate the associated conserved and thermodynamic quantities. The
four-dimensional action of Einstein-Maxwell gravity coupled to a dilaton
field is \cite{SHD1,SHD2} 
\begin{eqnarray}
\mathcal{S} &=&-\frac{1}{16\pi }\int d^{4}x\sqrt{-g}\Big(\mathcal{R}%
-2(\nabla \varphi )^{2}-V(\varphi )  \notag \\
&&-e^{-2\alpha \varphi }F_{\mu \nu }F^{\mu \nu }\Big),  \label{ac}
\end{eqnarray}%
where $\mathcal{R}$ is the Ricci scalar curvature, $\varphi $ is the dilaton
field and $V(\varphi )$ is the dilaton potential. Herein, the
electromagnetic field tensor $F_{\mu \nu }$ is defined in terms of the gauge
field $A_{\mu }$ via $F_{\mu \nu }=\partial _{\mu }A_{\nu }-\partial _{\nu
}A_{\mu }$. For an arbitrary value of the dilaton coupling strength $\alpha $
in AdS space, the dilaton potential is chosen to take the following form 
\cite{Gao,SH} 
\begin{eqnarray}
V(\varphi ) &=&\frac{2\Lambda }{3\left( \alpha ^{2}+1\right) ^{2}}\Big[%
8\alpha ^{2}e^{(\alpha ^{2}-1)\varphi /\alpha }-\left( \alpha ^{2}-3\right)
e^{2\alpha \varphi }  \notag \\
&&+\alpha ^{2}\left( 3\alpha ^{2}-1\right) e^{-2\varphi /\alpha }\Big],
\label{potential}
\end{eqnarray}%
where $\Lambda $ is the cosmological constant that relates to the AdS radius 
$l$ as $\Lambda =-3/l^{2}$. The potential given in Eq. (\ref{potential})
shows that the cosmological constant $\Lambda $ is coupled to the dilaton
field $\varphi $ in\ a non-trivial way. When the coupling constant $\alpha
=\pm 1/\sqrt{3},\pm 1,\pm \sqrt{3}$, the dilaton potential in Eq.(\ref%
{potential}) is indeed the SUSY potential of string theory. Note that in the
absence of the dilaton field, i.e. $V(\varphi =0)=2\Lambda $, the action Eq.(%
\ref{ac}) reduces to the usual Einstein-Maxwell theory with cosmological
constant. In $3+1$ dimensions, the line element of a static spherically
symmetric spacetime is written 
\begin{equation}
ds^{2}=-f(\rho )dt^{2}+\frac{d\rho ^{2}}{f(\rho )}+\rho ^{2}R^{2}(\rho
)d\Omega ^{2},
\end{equation}%
where $d\Omega ^{2}$ is the metric of the $2$-dimensional unit sphere with
volume $\omega =4\pi $ and the metric functions $f(\rho )$ and $R(\rho )$
are given by \cite{SHD1}%
\begin{equation}
f(\rho )=\left( 1-\frac{b}{\rho }\right) ^{\gamma }\left[ \left( 1-\frac{b}{%
\rho }\right) ^{1-2\gamma }\left( 1-\frac{c}{\rho }\right) +{\frac{\rho ^{2}%
}{l^{2}}}\right] ,  \label{metric}
\end{equation}%
\begin{equation}
R^{2}(\rho )=\left( 1-\frac{b}{\rho }\right) ^{\gamma },\text{\ }  \label{R}
\end{equation}%
where $b$ and $c$ are integration constants and $\gamma =2\alpha
^{2}/(\alpha ^{2}+1)$. Also, the dilaton field and the only non-vanishing
component of the gauge field $A_{\mu }$ are obtained as \cite{SHD1} 
\begin{equation}
\varphi (\rho )=\frac{\sqrt{\gamma \left( 2-\gamma \right) }}{2}\ln \left( 1-%
\frac{b}{\rho }\right) ,\quad A_{t}=-\frac{q}{\rho },
\end{equation}%
where $q$, an integration constant, is the charge parameter which is related
to $b$ and $c$ via the following relation 
\begin{equation}
q^{2}=\frac{bc}{\alpha ^{2}+1}.
\end{equation}%
For $\alpha \neq 0$, these solutions become imaginary in the range of $%
0<\rho <b$, so this region should be excluded from the spacetime. One may
also have a close look on the expansion of $V(\varphi )$. Given $\varphi
(\rho )$ at hand, it is a matter of calculation to show that for small $%
\alpha$, 
\begin{equation}
V(\varphi )=2\Lambda +4\Lambda \alpha ^{2}\Bigg{\{}\frac{b(\rho -7b/6)}{\rho
^{2}(1-b/\rho )^{2}}+\ln (1-b/\rho )\Bigg{\}}+\mathcal{O}(\alpha ^{4}),
\end{equation}%
which implies that, in the presence of dilaton field, the leading correction
term to the cosmological constant is of order $\alpha^2$. The black hole
event horizon is located at $\rho =\rho _{+}$ which is determined by the
largest real root of $f(\rho _{+})=0$. The mass and electric charge of the
dilaton AdS black hole per unit volume $\omega $ are \cite{SHD1}%
\begin{equation}
M=\frac{1}{8\pi }\left( c-b\frac{\alpha ^{2}-1}{\alpha ^{2}+1}\right) ,\text{
\ \ }Q=\frac{q}{4\pi }.
\end{equation}%
Also, the other associated thermodynamic quantities, such as the Hawking
temperature $T$, entropy $S$ and electric potential $U$, are 
\begin{eqnarray}
T &=&\frac{1}{4\pi \rho _{+}}\left( 1-\frac{b}{\rho _{+}}\right) ^{1-\gamma }%
\Bigg{\{}1+\frac{\rho _{+}}{l^{2}}\left[ 3\rho _{+}+2b\left( \gamma
-2\right) \right]  \notag \\
&&\times \left( 1-\frac{b}{\rho _{+}}\right) ^{2(\gamma -1)}\Bigg{\}},
\label{Tem}
\end{eqnarray}%
\begin{equation}
S=\frac{\rho _{+}^{2}}{4}\left( 1-\frac{b}{\rho _{+}}\right) ^{\gamma },%
\text{ \ \ \ \ }U=\frac{q}{\rho _{+}},  \label{En}
\end{equation}%
where entropy $S$ is written per unit volume $\omega $. It is easy to verify
that the first law of black hole thermodynamics 
\begin{equation}
dM=TdS+UdQ,
\end{equation}%
is satisfied on the event horizon \cite{SHD1}.

It is worthwhile to mention that in the absence of the dilaton filed ($%
\alpha =0$), the solutions reduce to the well-known four-dimensional
Reissner-Nordstrom (RN)-AdS black hole. It is also notable to mention that
these solutions are even functions in $\alpha $. In the next section, we
study the critical behavior of dilaton AdS black hole in the phase space. 
\begin{figure}[t]
\epsfxsize=8.5cm \centerline{\epsffile{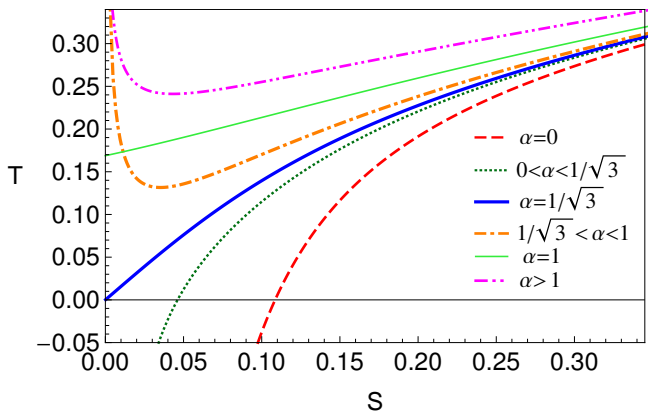}}
\caption{$T$-$S$ diagram of charged dilaton AdS black hole. This figure
shows the remarkable influence of the coupling constant $\protect\alpha $ on
the temperature. Here, we have set $l=1$ and $q=1$.}
\label{fig1}
\end{figure}
\begin{figure*}[t]
\begin{center}
\begin{minipage}[b]{0.46\textwidth}\begin{center}
                \subfigure[~$q_{c}$ versus $\alpha$]{
                    \label{fig2a}\includegraphics[width=\textwidth]{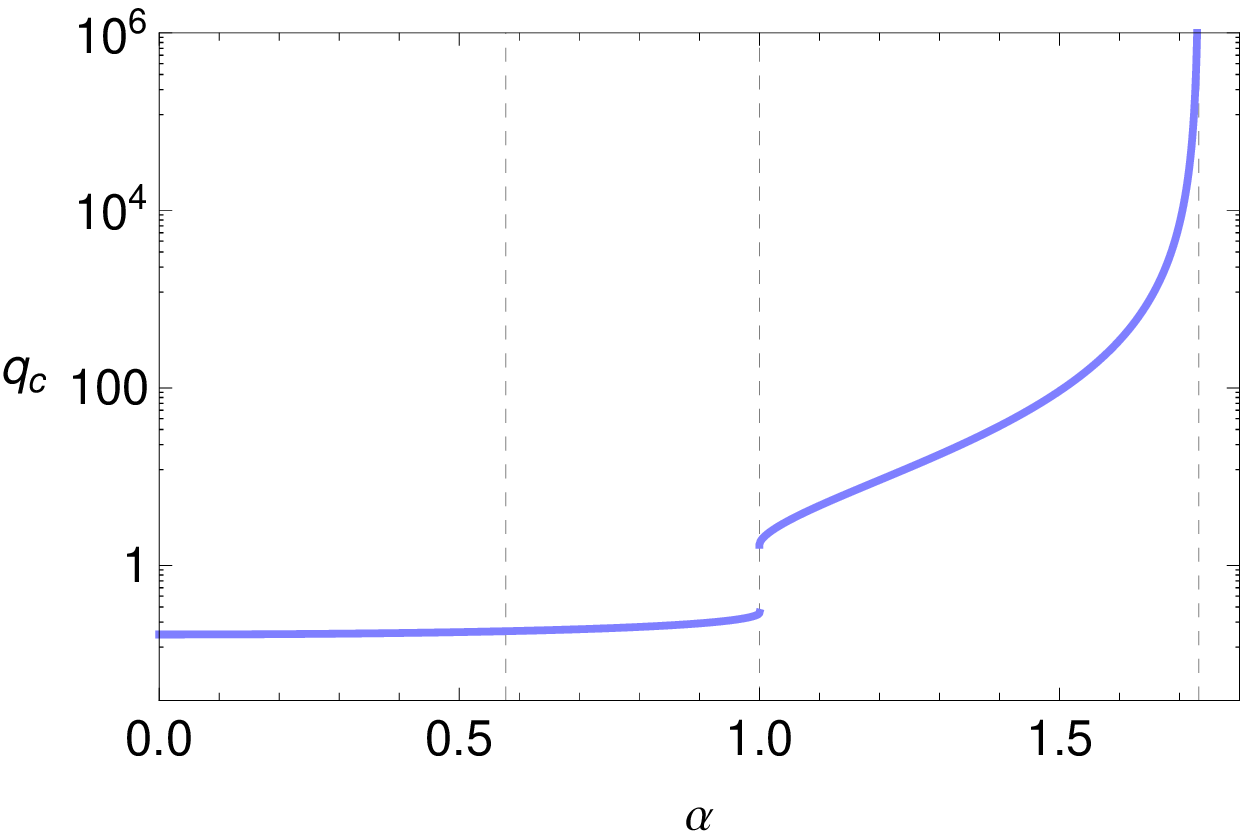}\qquad}
        \end{center}\end{minipage} \hskip+1cm 
\begin{minipage}[b]{0.46\textwidth}\begin{center}
                \subfigure[~$T_{c}$ versus $\alpha$]{
                    \label{fig2b}\includegraphics[width=\textwidth]{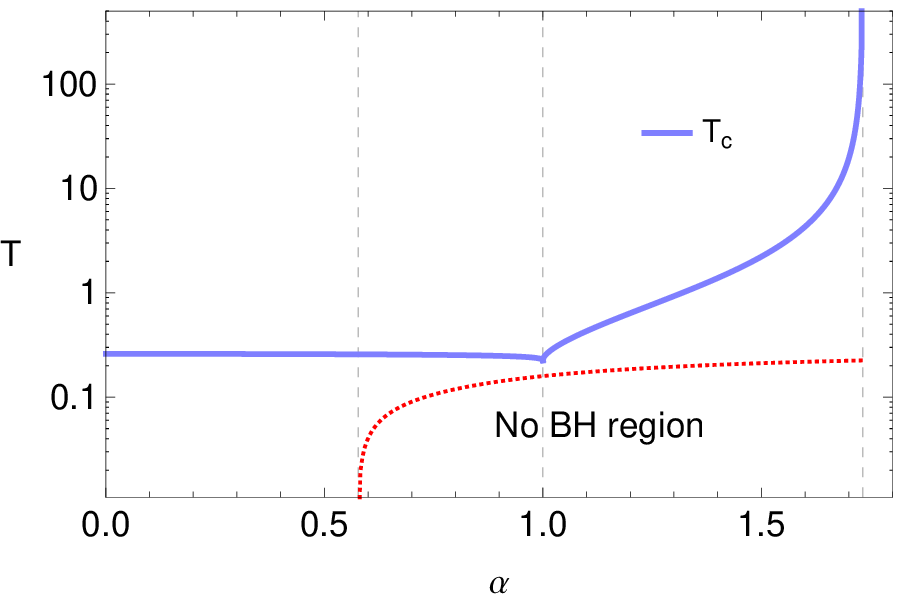}\qquad}
        \end{center}\end{minipage} \hskip0cm
\end{center}
\caption{The behaviors of the critical electric charge ($q_{c}$) and
critical temperature ($T_{c}$) versus $\protect\alpha $. The no BH region
corresponds to no BH solution. The vertical dashed lines mark the values of $%
\protect\alpha =1/\protect\sqrt{3}$, $\protect\alpha=1$ and $\protect\alpha=%
\protect\sqrt{3} $. We use the logarithmic scales on the vertical axis and
set $l=1$.}
\label{fig2}
\end{figure*}
\begin{figure*}[t]
\begin{center}
\begin{minipage}[b]{0.46\textwidth}\begin{center}
                \subfigure[~$\protect\rho _{+c}$ versus $\alpha$]{
                    \label{fig3a}\includegraphics[width=\textwidth]{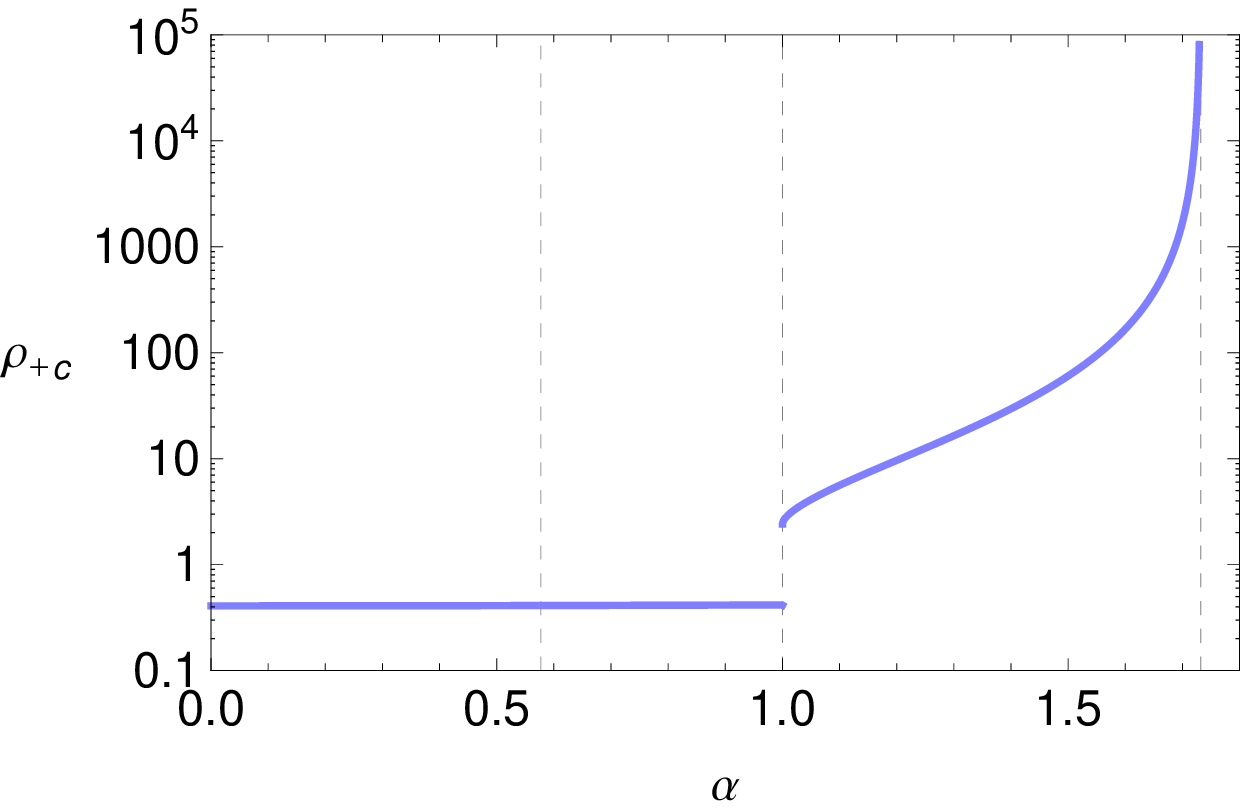}\qquad}
        \end{center}\end{minipage} \hskip+1cm 
\begin{minipage}[b]{0.46\textwidth}\begin{center}
                \subfigure[~$S_{c}$ versus $\alpha$]{
                    \label{fig3b}\includegraphics[width=\textwidth]{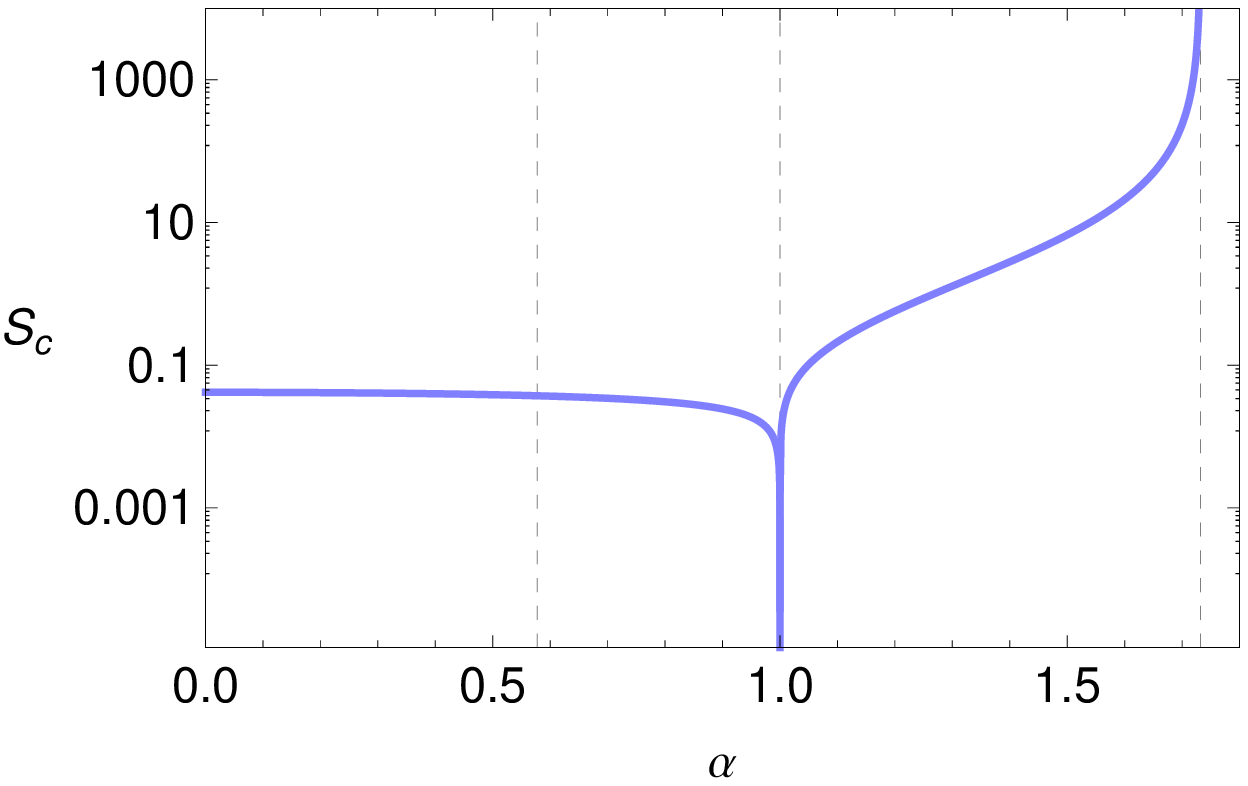}\qquad}
        \end{center}\end{minipage} \hskip0cm
\end{center}
\caption{The behaviors of the critical event horizon radius ($\protect\rho %
_{+c}$) and critical entropy ($S_{c}$) versus $\protect\alpha $. The
vertical dashed lines mark the values of $\protect\alpha =1/\protect\sqrt{3}$%
, $\protect\alpha=1$ and $\protect\alpha=\protect\sqrt{3}$. We use the
logarithmic scales on the vertical axis and set $l=1$.}
\label{fig3}
\end{figure*}

\section{Critical behavior of charged dilaton AdS black hole \label{Critical}%
}

In this section we are going to investigate the effects of the dilaton field
on the critical behavior of charged dilaton AdS black hole. To end this, we
analyze behavior of the specific heat at constant charge 
\begin{equation}
C_{q}=T\left( \frac{dS}{dT}\right) _{q},  \label{CQ}
\end{equation}%
where $l$ and $\alpha $ are also fixed. The sign of this quantity determines
the local thermodynamic stability, i.e. the stability (instability) is
accompanied by $C_{q}>0$ ($C_{q}<0$). To see the influence of the dilaton
field ($\alpha $) on $C_{q}$, we plot the behavior of the temperature as a
function of entropy in Fig. \ref{fig1} for different values of $\alpha $ and 
$q=1$. It is obvious from Fig. \ref{fig1}, that the behavior of the black
hole temperature significantly depends on $\alpha $ for small $S$.
Accordingly, we expand the temperature of the charged dilaton AdS black hole
for small entropy as follows:

\begin{itemizeReduced}
\item For $0<\alpha <1/\sqrt{3}\approx 0.58$,%
\begin{eqnarray}
T &=&\frac{\left( 3\alpha ^{2}-1\right) \left( \alpha ^{2}+1\right)
^{1/(\alpha ^{2}+1)-1}q^{2/(\alpha ^{2}+1)}}{\pi 2^{4/(\alpha
^{2}+1)+1}l^{2-2/(\alpha ^{2}+1)}S^{2/(\alpha ^{2}+1)-1/2}}  \notag \\
&&+\mathcal{O}\left( S^{2/(\alpha ^{2}+1)-5/2}\right) ,
\end{eqnarray}%
the black hole is \textquotedblleft \textit{Reissner-Nordstrom-AdS}%
\textquotedblright\ (RN) type in which with decreasing entropy, the
temperature goes over zero.

\item For $\alpha =1/\sqrt{3}$,%
\begin{eqnarray}
T &=&\frac{3\left( 3l^{2}+12q^{2}-l\sqrt{48q^{2}+9l^{2}}\right) }{2\pi
l^{4}q^{2}(\sqrt{9+48q^{2}/l^{2}}-3)^{3/2}}  \notag \\
&&\times \sqrt{3l^{2}+8q^{2}+l\sqrt{48q^{2}+9l^{2}}}S+\mathcal{O}\left(
S^{3}\right) ,
\end{eqnarray}%
where the dilaton black hole has zero temperature at the vanishing entropy
limit. This $\alpha $ may be called the \textquotedblleft \textit{marginal
coupling constant}\textquotedblright\ ($\alpha _{m}$).

\item For $1/\sqrt{3}<\alpha <1$,%
\begin{equation}
T=\frac{\left( 3\alpha ^{2}-1\right) \left( \alpha ^{2}+1\right)
^{1/(2\alpha ^{2})-1}}{\pi 2^{1/\alpha ^{2}+1}l^{2}q^{-1/\alpha
^{2}}S^{1/(2\alpha ^{2})-1/2}}+\mathcal{O}\left( S^{1/(2\alpha
^{2})-1/2}\right) ,
\end{equation}%
black hole is \textquotedblleft \textit{Schwarzschild-AdS}%
\textquotedblright\ \textbf{(}Schw\textbf{)}-type. In this case, black hole
solution does not exist in the low-temperature regime.

\item For $\alpha =1$,%
\begin{equation}
T=\frac{l^{2}+2q^{2}}{4\sqrt{2}l^{2}\pi q}+\mathcal{O}\left( S\right) ,
\end{equation}%
which is the \textquotedblleft \textit{spacial}\textquotedblright\ case
where the dilaton black hole has finite temperature at $S=0$.

\item For $\alpha >1$,%
\begin{equation}
T=\frac{2^{1/\alpha ^{2}-3}\left( \alpha ^{2}+1\right) ^{-1/(2\alpha ^{2})}}{%
\pi q^{1/(\alpha ^{2})}S^{1/2-1/(2\alpha ^{2})}}+\mathcal{O}\left(
S^{1/2-1/(2\alpha ^{2})}\right) ,
\end{equation}%
black hole is Schw-type again. As can be seen from Fig. \ref{fig1}, the
right branch of isocharge for Schw-type black hole is locally stable, i.e.,
the specific heat at constant charge is positive. On the other hand, the
large entropy limit of the temperature is 
\begin{equation}
T\approx 3\frac{\sqrt{S}}{2\pi l^{2}}\ \Rightarrow \ \ C_{q}=2S>0,
\end{equation}%
which is independent of the charge and dilaton coupling constant and always
yields a thermal stable system.
\end{itemizeReduced}

\begin{figure*}[t]
\begin{center}
\begin{minipage}[b]{0.46\textwidth}\begin{center}
                \subfigure[~{$\alpha =0.5 \in [0,1/\sqrt{3}]$}]{
                    \label{fig4a}\includegraphics[width=\textwidth]{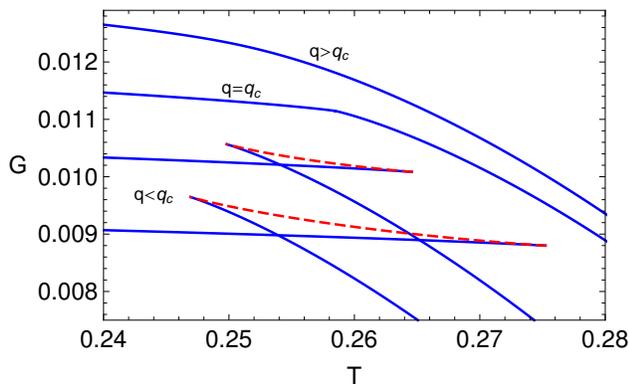}\qquad}
        \end{center}\end{minipage} \hskip+1cm 
\begin{minipage}[b]{0.46\textwidth}\begin{center}
                \subfigure[~$\alpha =0.7 \in
                ( 1/\sqrt{3},1) $]{
                    \label{fig4b}\includegraphics[width=\textwidth]{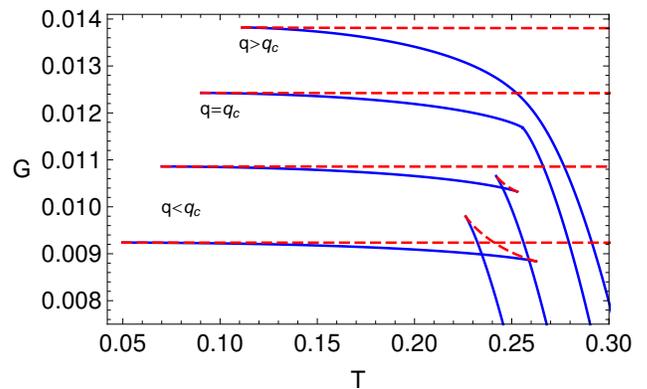}\qquad}
        \end{center}\end{minipage} \hskip0cm
\end{center}
\caption{Gibbs free energy as a function of temperature for $l=1$ and
various values of $q$. For $q<q_{c}$, the system undergoes a first order
phase transition between SBH and LBH. The positive (negative) sign of $C_{q}$
is identified by the blue solid (dashed red) line. The curves are shifted
for clarity.}
\label{fig4}
\end{figure*}
\begin{figure}[t]
\epsfxsize=8.0cm \centerline{\epsffile{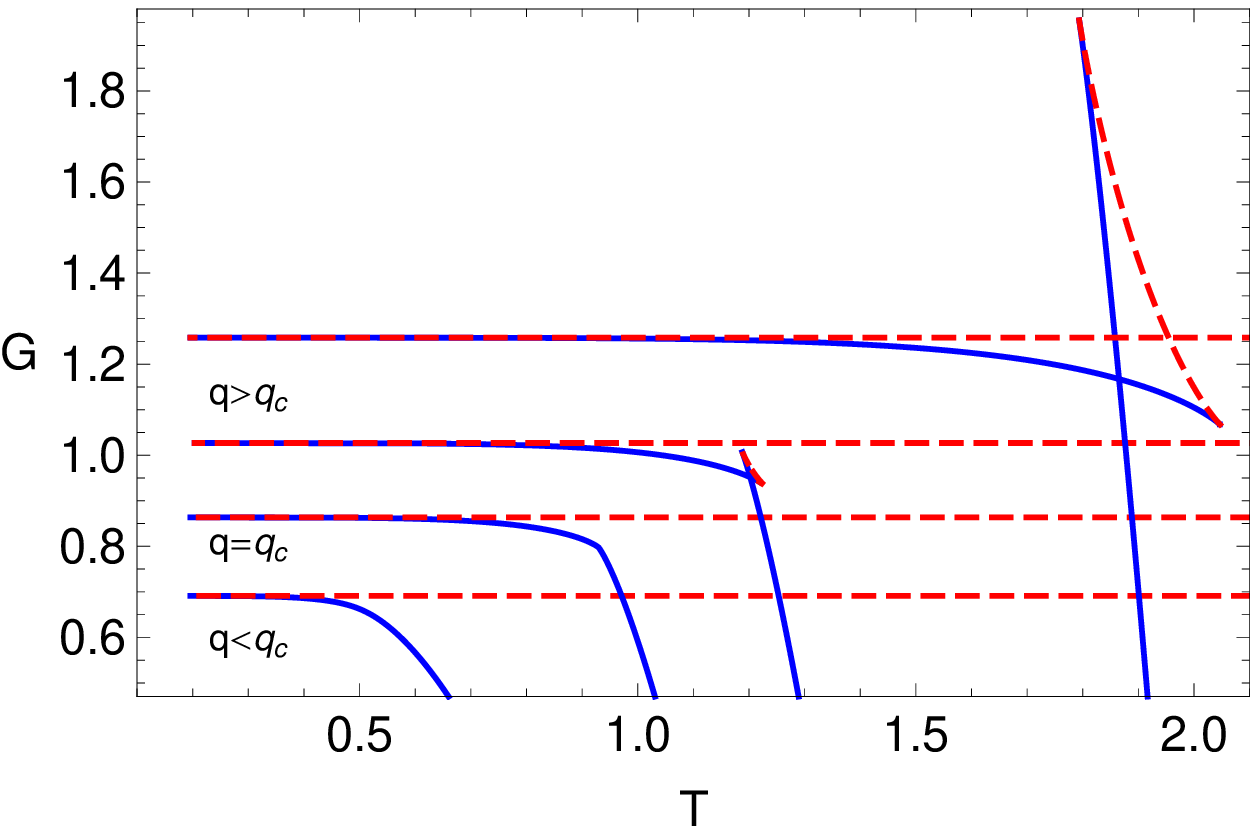}}
\caption{Gibbs free energy as a function of temperature for $l=1$, $\protect%
\alpha =1.3\in \left(1,\protect\sqrt{3}\right)$ and various values of $q$.
For $q>q_{c}$, the system undergoes a first order phase transition between
SBH and LBH. The positive (negative) sign of $C_{q}$ is identified by the
blue solid (dashed red) line. The curves are shifted for clarity.}
\label{fig5}
\end{figure}
\begin{figure*}[t]
\begin{center}
\begin{minipage}[b]{0.46\textwidth}\begin{center}
                \subfigure[]{
                    \label{fig6a}\includegraphics[width=\textwidth]{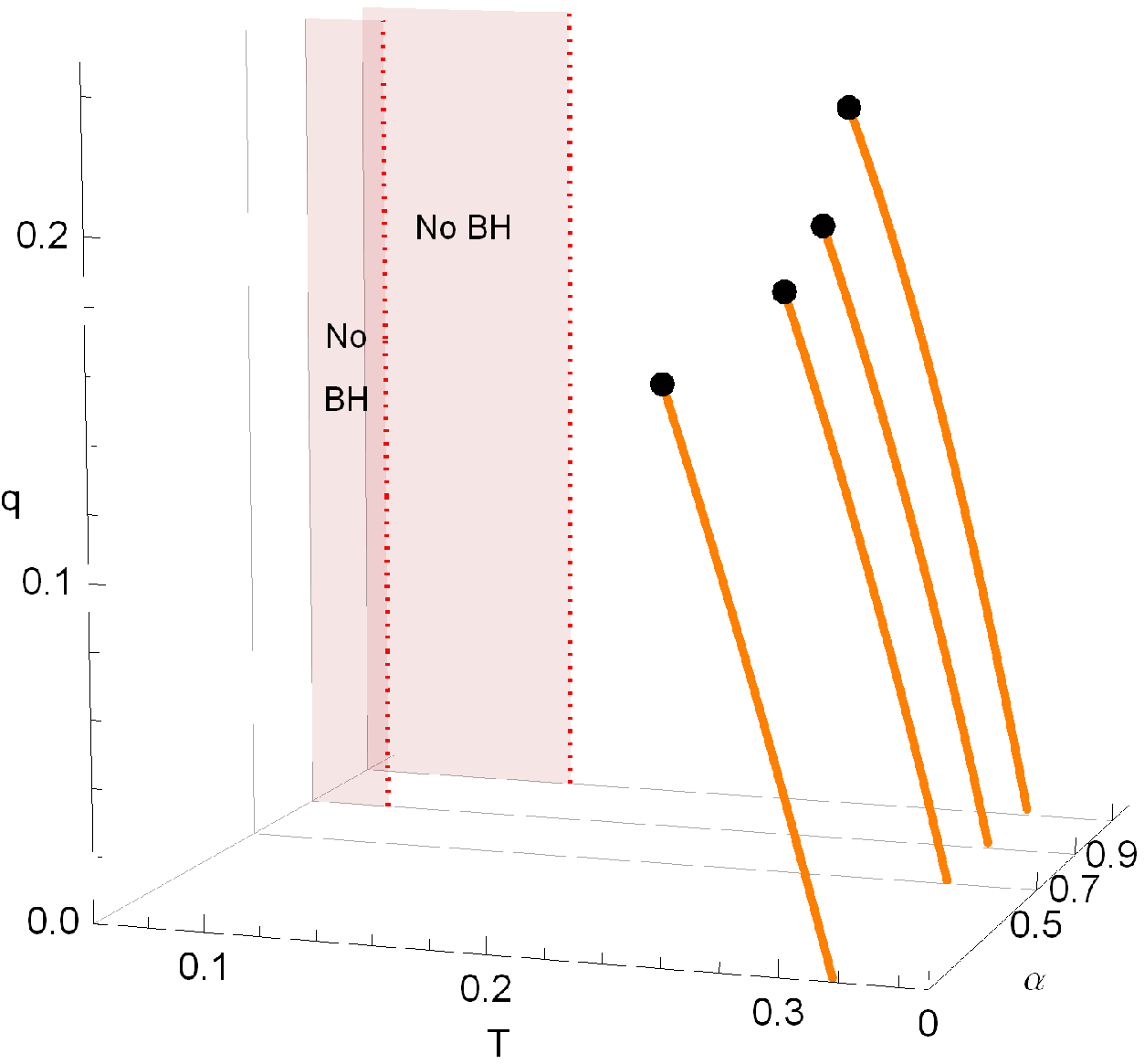}\qquad}
        \end{center}\end{minipage} \hskip+1cm 
\begin{minipage}[b]{0.39\textwidth}\begin{center}
                \subfigure[]{
                    \label{fig6b}\includegraphics[width=\textwidth]{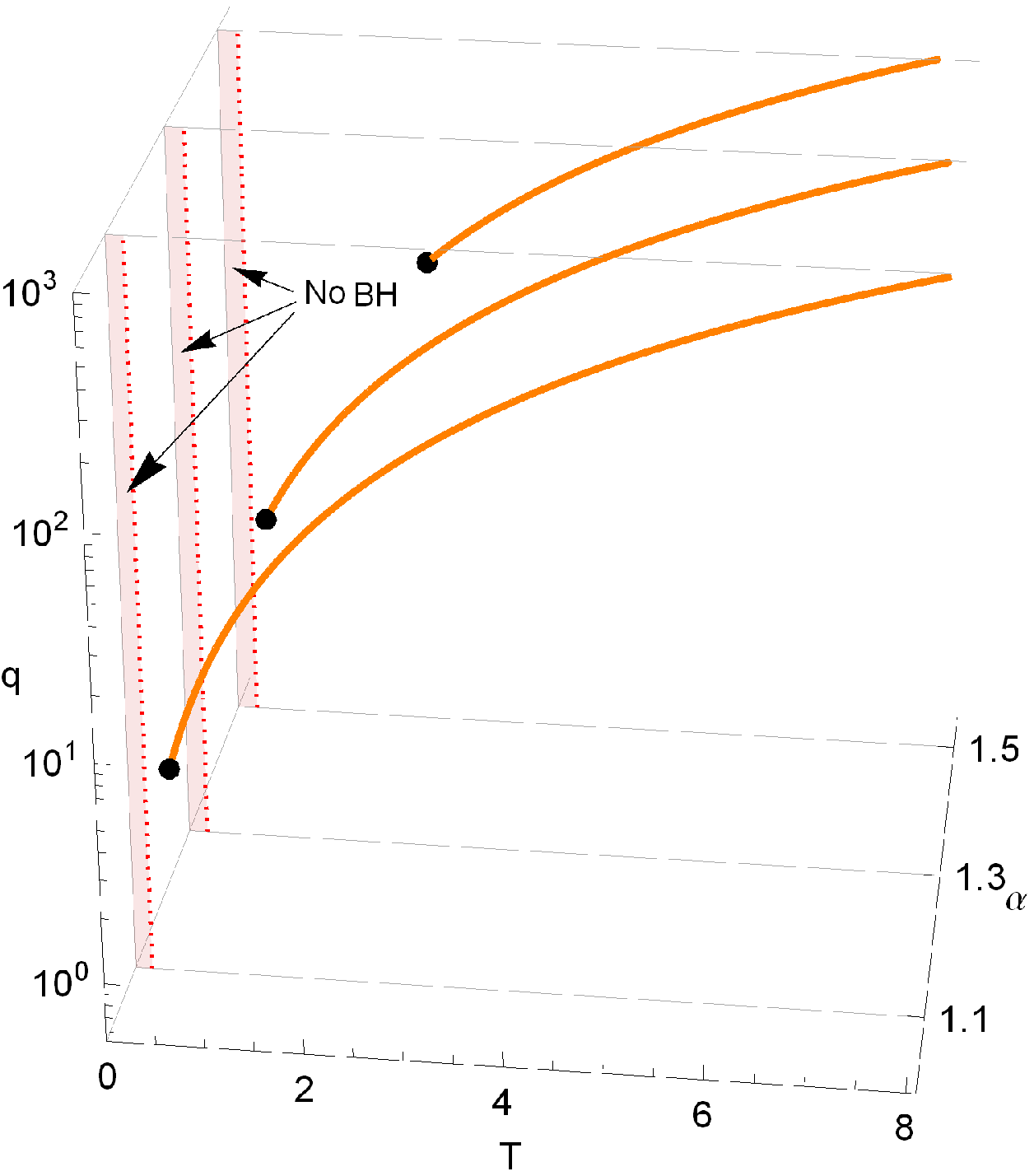}\qquad}
        \end{center}\end{minipage} \hskip0cm
\end{center}
\caption{SBH/LBH phase diagram for $l=1$ and various values of $\protect%
\alpha $. The critical points and first order phase transition curves are
highlighted by the black solid circle and solid orange line, respectively.
At low temperature, no BH regions correspond to no black hole solution. We
use the logarithmic scale on $q$ axis in Fig. (b).}
\label{fig6}
\end{figure*}
In what follows, we are going to obtain the critical point, which
corresponds to a second order phase transition, for various type of charged
dilaton AdS black holes. For fixed $q$ and $l$, the value of the critical
point is characterized by the inflection point 
\begin{equation}
\frac{\partial T}{\partial S}\Big|_{q_{c}}=0,\quad \quad \quad \frac{%
\partial ^{2}T}{\partial S^{2}}\Big|_{q_{c}}=0.  \label{cpoint}
\end{equation}%
To calculate the above expressions from Eq. (\ref{Tem}) and (\ref{En}), the
following relation is used 
\begin{equation}
\left. \frac{\partial T}{\partial S}\right\vert _{q}=\frac{\left. \frac{%
\partial T}{\partial \rho _{+}}\right\vert _{b,q}+\left. \frac{\partial T}{%
\partial b}\right\vert _{\rho _{+},q}\left. \frac{\partial b}{\partial \rho
_{+}}\right\vert _{q}}{\left. \frac{\partial S}{\partial \rho _{+}}%
\right\vert _{b}+\left. \frac{\partial S}{\partial b}\right\vert _{\rho
_{+}}\left. \frac{\partial b}{\partial \rho _{+}}\right\vert _{q}},  \notag
\end{equation}%
where 
\begin{equation}
\left. \frac{\partial b}{\partial \rho _{+}}\right\vert _{q}=-\frac{\left. 
\frac{\partial f(\rho _{+})}{\partial \rho _{+}}\right\vert _{b,q}}{\left. 
\frac{\partial f(\rho _{+})}{\partial b}\right\vert _{\rho _{+},q}}.  \notag
\end{equation}%
Due to the complicated form of Eqs. (\ref{Tem}) and (\ref{cpoint}), it is
almost impossible to obtain the critical values, analytically. Hence, we
numerically solve the set of Eq. (\ref{cpoint}) for a given value of $\alpha 
$. The calculated values of the critical quantities, such as $q_{c}$, $T_{c}$%
, $\rho _{+c}$ and $S_{c}$, for various $\alpha $ are illustrated in Figs. %
\ref{fig2} and \ref{fig3}. We observe that there is no critical point for
charged dilaton AdS black hole in cases where $\alpha =1$ and $\alpha \geq 
\sqrt{3}\approx 1.73$. As one can see from Figs. \ref{fig2} and \ref{fig3},
for $0<\alpha <1$, the dilaton coupling parameter ($\alpha $) does not
significantly affect the critical quantities, except entropy which abruptly
decreases close to $1$. Fig. \ref{fig2b} shows that the critical point
occurs in the RN-type of black hole when $0\leq \alpha <1/\sqrt{3}$, whereas
for $1/\sqrt{3}<\alpha <1$ it occurs in Schw-type where there is a lower
bound on temperature of the black hole. As expected from Figs. \ref{fig2}
and \ref{fig3}, in the absence of dilaton field \ ($\alpha =0$), the
critical quantities reduce to those of charged AdS black hole \cite{PV}. In
case of $1<\alpha <\sqrt{3}$, with increasing $\alpha $, the values of
critical quantities increase and diverge for $\alpha \rightarrow \sqrt{3}$.
It should also be pointed out that in the range $1<\alpha <\sqrt{3}$, the
critical behavior happens in Schw-type region.

In order to fully obtain phase transition and examine phase structure of
charged dilaton AdS black holes, we shall study the behavior of Gibbs free
energy in the next section.


\section{Gibbs free energy \label{Gibbs}}

The general thermodynamic description of charged dilaton AdS black hole is
provided by studying the Gibbs free energy which exhibits the global stable
state. The Gibbs free energy for a fixed AdS radius regime can be obtained
through the Legendre transformation of the mass $M$. Thus, the Gibbs free
energy per unit volume $\omega $ is given 
\begin{eqnarray}
G\left( T,q\right) &=&M-TS \\
&=&\frac{l\left( \Upsilon \left[ 3-4\Gamma -\alpha ^{2}\right] +2\left(
\alpha ^{2}-1\right) +\alpha ^{2}+5\right) }{32\sqrt{2}\pi \left( \alpha
^{2}+1\right) \Gamma ^{3/2-2/(\alpha ^{2}+1)}\left( \Upsilon -1\right)
^{-1/2}},  \notag
\end{eqnarray}%
where 
\begin{equation}
\Upsilon \equiv \sqrt{1+\frac{4q^{2}\left( \alpha ^{2}+1\right) \Gamma
^{3-4/(\alpha ^{2}+1)}}{l^{2}\left( 1-\Gamma \right) }},  \notag
\end{equation}%
and $\Gamma =1-b/\rho _{+}$, thus we have $\Gamma =\Gamma \left( T,q\right) $%
. Notice that the reality condition of the black hole solutions ($\rho>b$)
leads to the constraint $0<\Gamma <1$.

The behavior of Gibbs free energy in terms of the temperature $T$ for $%
\alpha =0.5$, $0.7$ and $1.3$ are depicted in Figs. \ref{fig4} and \ref{fig5}%
, for different values of charge $q$. From these figures, it can be seen
that the charge dependence of Gibbs free energy is strongly affected by the
dilaton coupling parameter $\alpha $. In case of $\alpha =0.5\in \left[ 0,1/%
\sqrt{3}\right] $, where charged dilaton AdS black hole is RN-type, the
Gibbs free energy is single value in temperature for $q>q_{c}$ (see Fig. \ref%
{fig4a}). In this case, back hole is locally stable ($C_{q}>0$) which is
indicated by the solid blue line in Fig. \ref{fig4a}. On the other hand,
when $q<q_{c}$, black hole becomes thermodynamically unstable where the
Gibbs free energy is multi-valued in the certain range of temperature. This
corresponds to $C_{q}<0$ which is shown by dashed-red line in Fig. \ref%
{fig4a}. This unstable behavior of the Gibbs free energy for $q<q_{c}$
exhibits a \textit{first order} (discontinuous) phase transition between
small black hole (SBH) and large black hole (LBH). For $\alpha =0.7$ $\in
\left( 1/\sqrt{3},1\right) $ case, as illustrated in Fig. \ref{fig4b},
charged dilaton AdS black hole is Schw-type where the lower (upper) branch
of the Gibbs free energy is globally stable (unstable) for $q>q_{c}$. For $%
q<q_{c}$, a first order phase transition occurs between SBH and LBH in the
lower branch of the Gibbs free energy which is stable. For $\alpha =1.3$ $%
\in \left( 1,\sqrt{3}\right) $, a novel behavior happens in for Schw-type
black hole (see Fig. \ref{fig5}). Indeed, in contrast to what occurs in Fig. %
\ref{fig4b}, in this case a first order\ phase transition takes place
between SBH and LBH for $q>q_{c}$. This behavior has not been observed in
previous studies on phase transition of charged AdS black holes \cite%
{AAA,ASheykhi}. It is notable to mention that we do not find any other phase
transition for charged dilaton AdS black hole.

\bigskip The corresponding SBH/LBH phase diagram of dilaton AdS black hole
for different values of the dilaton parameter $\alpha $ is sketched in Fig. %
\ref{fig6}. It is clear from Fig. \ref{fig6} that the critical points are
denoted by black spots at the end of the first order phase transition curves
(orange). In Fig. \ref{fig6a}, the first order phase transition curves
separate the SBH from LBH for $q<q_{c}$, while in Fig. \ref{fig6b}, these
curves distinguish the SBH from LBH for $q>q_{c}$. Also, no BH regions
implies that no BH solutions exist at the low temperature. 

\section{Summary \label{Summary}}

To sum up, we have revisited critical behaviour and phase structure of
charged dilaton black holes in the background of AdS spaces. The motivation
for study phase behaviour of dilaton black holes in AdS spacetime is mainly
inspired by AdS/CFT correspondence and is expected to shed light on the
microscopic structure of black holes. We adopted the view point that
cosmological constant can be regarded as a fixed parameter, while the charge
of the black hole varies.

To understand the impact of the dilaton field on the heat capacity, $C_{q}$,
which determines the local thermodynamic stability of the system, we have
studied the behavior of the temperature $T$ as a function of entropy $S$ for
different values of $\alpha $. By expanding $T$ for small values of $S$, we
have distinguished several black hole systems, with thermal
stability/instability, depending on the values of $\alpha$. In order to
obtain the coordinates of the critical point, we numerically solved the
system of equations and plotted the quantities at the critical point in
terms of $\alpha $. We observed that there is no critical point in cases
with $\alpha =1$ and $\alpha \geq \sqrt{3}$, while for $0<\alpha <1$, the
critical quantities are not significantly affected by $\alpha$, except
entropy which abruptly decreases close to $1$. In the absence of dilaton
field ($\alpha =0$), the critical quantities reduce to those of charged AdS
black hole.

We have also studied the Gibss free energy, which exhibits the global stable
state of the system, for different values of $\alpha$ and $q$. We have
realized several cases, depending on $\alpha$, including whether or not the
Gibbs free energy is single/multi-valued and whether or not the system is
thermally stable/unstable. Interestingly enough, we realized that unstable
behavior of the Gibbs free energy for $q<q_{c}$ exhibits a \textit{first
order} (discontinuous) phase transition between SBH and LBH for $%
0\leq\alpha<1$. For $1<\alpha<\sqrt{3}$, however, a novel first
order phase transition happens between SBH and LBH provided $q>q_{c}$. The
later has not been observed in the previous studies on phase transition of
charged AdS black holes and is one of the new result of the present paper.


\begin{acknowledgments}
We are grateful to the Research Council of Shiraz University. AS thanks
Herman Nicolai and Max-Planck-Institute for Gravitational Physics, for
hospitality.
\end{acknowledgments}


\end{document}